\documentclass[9pt]{article}
\usepackage{spconf,amsmath,graphicx,hyperref}
\usepackage{algorithm}
\usepackage{algpseudocode}
\usepackage{xcolor}

\title{FINITE SCALAR QUANTIZATION ENABLES REDUNDANT AND TRANSMISSION-ROBUST NEURAL AUDIO COMPRESSION AT LOW BIT-RATES}
\name{Harry Julian, Rachel Beeson, Lohith Konathala, Johanna Ulin, Jiameng Gao}
\address{Neuphonic}

\begin{document}
\ninept

\maketitle
\begin{abstract}
Neural Audio Codecs (NACs) have become increasingly adopted in speech processing tasks due to their excellent rate-distortion performance and compatibility with Large Language Models (LLMs) as discrete feature representations for audio generation. While most existing codecs rely on Residual Vector Quantization (RVQ), Finite Scalar Quantization (FSQ) has recently emerged as a compelling alternative that simplifies training and natively supports single codebooks. We introduce NeuCodec, an FSQ-based NAC, and show that FSQ encodes baked-in redundancy which produces an encoding which is robust when transmitted through noisy channels. First, through an encoder distillation experiment, we show that two different encoders can learn to encode identical audio into vastly different code sequences whilst maintaining comparable reconstruction quality with the same quantizer and decoder. Second, we demonstrate that FSQ has vastly superior bit-level perturbation robustness by comparing the performance of RVQ and FSQ codecs when simulating the transmission of code sequences through a noisy channel. 

\end{abstract}
\begin{keywords}Audio Compression, Neural Compression, Neural Audio Codec, Residual Vector Quantization, Finite Scalar Quantization.\end{keywords}

\section{Introduction}
\label{sec:intro}

Recently, Neural Audio Codecs (NACs) have gained widespread usage in speech processing, due to their ability to compress speech into ultra-low bitrate discrete code sequences whilst maintaining high perceptual quality when reconstructing these sequences back into waveforms \cite{guo2025recentadvancesdiscretespeech}.

The autoencoding task used to train NACs embeds a compressed latent representation of speech features into discrete sequences of codes, which are useful for training autoregressive transformers to complete downstream audio tasks such as Text-to-Speech (TTS) \cite{lyth2024naturallanguageguidancehighfidelity}, Automatic Speech Recognition (ASR) \cite{Dhawan_2024} and Full Duplex Speech Modeling \cite{défossez2024moshispeechtextfoundationmodel}; they can also be used as a domain-specific tokenized vocabulary that Large Language Models (LLMs) can be adapted to use for audio generation \cite{ye2025llasascalingtraintimeinferencetime}.

Conventionally, the most widely used NACs have utilized Residual Vector Quantization (RVQ) \cite{zeghidour2021soundstreamendtoendneuralaudio}, where at each encoder output timestep, the encoded feature representation is quantized by a top-level `coarse' codebook, and additional codebooks quantize the residual error from each prior quantization operation. Although effective, RVQ presents training challenges, as propagating gradients to the codeword vectors to align them with the unquantized encoder outputs necessitates the use of auxiliary loss functions. This creates a delicate optimization problem that often leads to codebook collapse \cite{kumar2023highfidelityaudiocompressionimproved} where only a subset of codewords is used. Additionally, RVQ also requires a comparatively complicated downstream modeling setup, as the sequence length is expanded by the number of quantized residuals; mechanisms to model the hierarchical nature of RVQ codes commonly rely on two separate transformers that operate globally and locally \cite{wang2023neuralcodeclanguagemodels}.

Finite Scalar Quantization (FSQ) \cite{mentzer2023finitescalarquantizationvqvae}, a method that uses a simple fixed-grid for partitioning the codebook, constructs a single codebook by quantizing each output vector dimension, treating each dimension as an implicit codebook, rather than quantizing an entire latent vector as a whole. Using FSQ results in almost complete codebook utilization, requires no auxiliary losses to train and affords simpler downstream architectures due to the usage of a single codebook, rather than multiple recursively dependent codes needing to be predicted per timestep.

Through experimentation with our codec, NeuCodec, we show that FSQ-based codecs also exhibit an additional perturbation robustness property in their code sequences. First, we introduce NeuCodec, our FSQ-based codec model. Second, via an encoder distillation experiment with NeuCodec, we show that two encoders can learn to encode the same audio in very different code sequences given a fixed quantizer and decoder, yet the sequence can be reconstructed to a similar perceptual fidelity from both sequences; analyzing the differences between the representations suggests the learned encoding is localized and has redundancy baked-in. Third, via a perturbation experiment where we simulate transmission of codes from various FSQ and RVQ codecs through a noisy channel, we show that FSQ-based codecs exhibit better performance under reasonably large levels of perturbation. We offer explanations for this phenomenon and speculate on future applications of FSQ-based codecs in light of this property.

\section{Background}
\label{sec:Background}

RVQ discretizes an embedding space through first performing Vector Quantization \cite{oord2018neuraldiscreterepresentationlearning} over a finite codebook, after which discretization errors (e.g. the distance between the scalar vector and the nearest neighbor codeword embedding) are obtained and discretized again, a process that continues for a predetermined number of codebooks. This means that scalar embeddings can be accurately represented through a hierarchical sequence of discretized tokens, all contained within a finite vocabulary.

FSQ creates discretized tokens from a continuous scalar space by projecting the latent space of the encoder space down to a much lower dimension and quantizing each dimension in the space to a number of scalar levels. Tokens are then obtained by enumerating through the discretized levels in each dimension. The encoder output is projected into a space where each dimension `d` is bounded between $[-1, 1]$ and then discretized to one of $n$ equidistant values. The codebook size $C$ is given by Eq.~\ref{eq:codebook}.

\begin{equation}
C = \prod_{i=1}^{d} n_{i}
\label{eq:codebook}
\end{equation}

Importantly, this implies that the output of the encoder is projected and quantized into a vector that can be mapped to a discrete set of values. Therefore, codebooks of the same size with the same $n$ values for each dimension will result in the same partitioning of the bounded quantization space. As a decoder operates on this quantization space, it means that two encoders that learn a similar partitioning of the fixed quantization space could utilize the same decoders without retraining.

For the experiment in Section~\ref{sec:bit_perturb} we make comparisons between our codecs and other NACs. For RVQ, we use Encodec \cite{défossez2022highfidelityneuralaudio} and Descript Audio Codec (DAC) \cite{kumar2023highfidelityaudiocompressionimproved}, both of which are mainly composed of convolutions. For FSQ, we use our own models as well as Stable Codec \cite{parker2024scalingtransformerslowbitratehighquality}, a large transformer based codec with 1B parameters.

\section{NeuCodec}

NeuCodec is primarily based on XCodec2 \cite{ye2025llasascalingtraintimeinferencetime}, an ultra-low bitrate audio codec designed for downstream modeling in LLM-based TTS. The encoder takes raw waveforms as an input and consists of a pre-trained frozen semantic encoder and a trainable acoustic encoder. The semantic encoder is Wav2Vec2-BERT-large \cite{communication2023seamlessmultilingualexpressivestreaming} which was pre-trained on 4.5 million hours of unsupervised speech. The acoustic encoder is derived from the encoder of BigCodec \cite{xin2024bigcodecpushinglimitslowbitrate} which is a stack of Residual CNNs with Snake activation functions \cite{ziyin2020neuralnetworksfaillearn}. The discrete bottleneck of the codec is an FSQ module with a projection dimension of $8$ and a codebook size of $2^{16}$. The decoder is a standard transformer decoder, which is used to directly predict magnitude and phase for a Vocos \cite{siuzdak2023vocos} head that generates a waveform.

The base model was trained for 800k steps following the approach of XCodec2 on one 8xH100 node with an effective batch size of 96 across GPUs. During training, each batch item is randomly cut into a 6 second segment (or padded if shorter). Training data are described in Table~\ref{tab:datasets}. The datasets used were selected as they are licensed for commercial usage (in contrast to the original XCodec2). Evaluation of all trained models is presented in Table~\ref{tab:model_performance}.

Additionally, we froze the weights of the encoder and quantizer and trained a new 24kHz upsampling decoder by increasing the hop-length from 320 to 480 to enable 16kHz to 24kHz upsampling. The model was trained for 200k steps using the same compute configuration. A 24khz subset of the data was used to train the upsampling decoder.

\begin{table}[ht]
  \centering
  \caption{NeuCodec Training Data Sources.}
  \label{tab:datasets}
  \begin{tabular}{lrlc}
    \hline
    \textbf{Dataset} & \textbf{Hours} & \textbf{Subset} & \textbf{License} \\
    \hline
    Emilia-YODAS \cite{he2025emilialargescaleextensivemultilingual} & 110{,}000 & 16 & CC-BY \\
    MLS \cite{Pratap_2020} & 45{,}000 & 16 & CC-BY \\
    LibriTTS-R \cite{koizumi2023librittsrrestoredmultispeakertexttospeech} & 585 & 16/24 & CC-BY \\
    Fleurs-R \cite{ma2024fleursrrestoredmultilingualspeech} & 692 & 16/24 & CC-BY \\
    Common Voice Subset \cite{ardila2020commonvoicemassivelymultilingualspeech} & 9{,}283 & 16 & CC0 \\
    HUI \cite{puchtler2021huiaudiocorpusgermanhighqualitytts} & 326 & 16/24 & CC0 \\
    Proprietary & 1000 & 16/24 & --- \\ 
    \hline
    \textbf{Total 16kHz} & \textbf{166{,}930} & & \\
    \textbf{Total 24kHz} & \textbf{2{,}603} & & \\
    \hline
  \end{tabular}
\end{table}

\label{sssec:dataset_overview}
\vspace{-0.5cm}
\section{Encoder Distillation}
\label{sec:encoder_distillation}

XCodec2 was originally designed as a feature representation for TTS. Its asymmetric configuration of an encoder that largely outsizes the decoder in parameter size and compute complexity, enables a trade-off of enhanced compression performance and slow encoding speeds with fast decoding speeds at inference time. As a trained TTS model is decode-heavy, this offsets most of the computation to training time where code sequences need to be generated beforehand. We set out to distill NeuCodec for low latency usage in encode-heavy paradigms (e.g. ASR).

\subsection{Training}

We modify the encoder architecture of NeuCodec, whilst mirroring the joint semantic and acoustic encoder paradigm used in the original model. We swap the BigCodec acoustic encoder with the L3AC Encoder \cite{zhai2025l3aclightweightlosslessaudio} (60\% of the original size) and swap Wav2VecBERT2.0 with DistillHubert \cite{chang2022distilhubertspeechrepresentationlearning} (4\% of the original size). Although the change in parameter count for the acoustic encoder is modest, it is mainly motivated by the fact that the BigCodec encoder has an abnormally high ratio of Multiply-Accumulate Operations (MACs) to parameters due to its deep stacking of convolutions \cite{wu2025ts3codectransformerbasedsimplestreaming}.

\begin{figure}[h]
  \centering
  \includegraphics[width=0.45\textwidth]{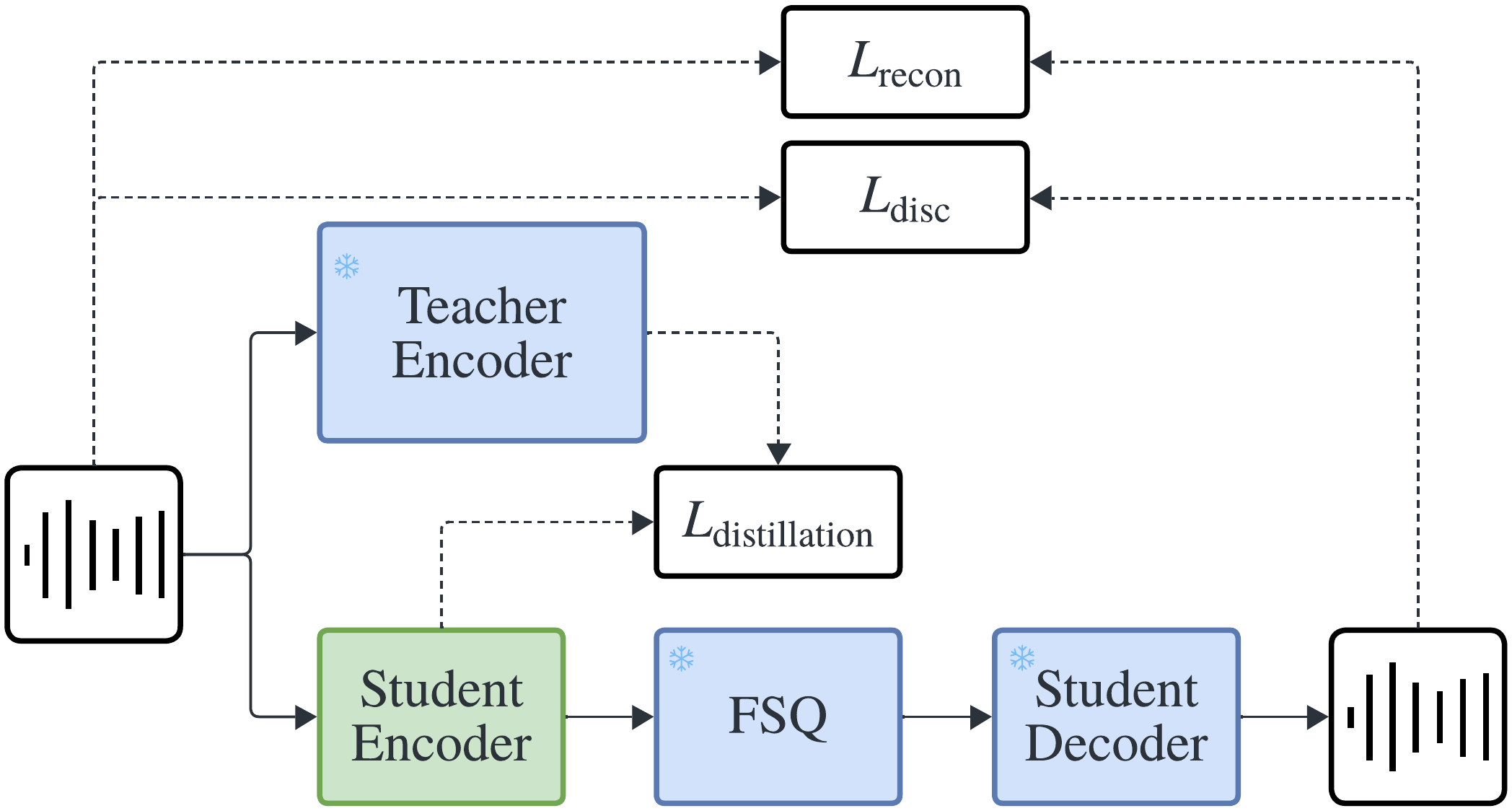}
  \caption{Distillation Training. Blocks labelled with snowflakes are frozen during training, with the remaining non-loss blocks being trained.}
  \label{fig:distillation_architecture}
\end{figure}

For distillation training, we add the distillation loss in Eq.~\ref{eq:distillation_loss} to push the encoded representations of the teacher and student encoders to be more similar: 

\begin{equation}
\mathcal{L}_{\text{distillation}} = \text{MSE}(\mathbf{h}_{\text{teacher}}, \mathbf{h}_{\text{student}})
\label{eq:distillation_loss}
\end{equation}

where $\mathbf{h}$ represents the pre-quantization encoder outputs of each respective encoder. This approach requires the output shapes of the student encoder to be the same as those of the teacher. 

The distillation loss is added to the original loss function used to train XCodec2, which is comprised of a multi-resolution mel-spectrogram loss \cite{défossez2022highfidelityneuralaudio} an average of the losses predicted by the Spectrogram Discriminator \cite{parker2024scalingtransformerslowbitratehighquality} and the HiFiGAN multi-period discriminator \cite{kong2020hifigangenerativeadversarialnetworks}, a discriminator feature matching loss and an L2 semantic reconstruction loss to make the final loss eq \ref{eq:losses}:

\begin{equation}
\begin{split}
\mathcal{L}_{\text{Total}} =\;& \lambda_{1} \mathcal{L}_{\text{mel-spec}}
+ \lambda_{2} \mathcal{L}_{\text{disc}}
+ \lambda_{3} \mathcal{L}_{\text{fm}} \\[6pt]
&+ \lambda_{4} \mathcal{L}_{\text{semantic}}
+ \lambda_{5} \mathcal{L}_{\text{distillation}}
\end{split}
\label{eq:losses}
\end{equation}

In training, the weights of the FSQ bottleneck and decoder are frozen. We use the 16kHz decoder to train the model, as it allows for use of a far larger pool of data. We train the student model for 400k steps on a single 8xH100 node with an effective batch size of 192, using the same dataset as NeuCodec. The distillation loss was activated after 20k steps, as activation at the beginning of training led the model to diverge due to large initial magnitudes of the loss. 

\begin{table*}
  \centering
  \caption{Encoder/decoder parameter breakdown and performance comparison on CMU-Arctic subset.}
  \label{tab:model_performance}
  \begin{tabular}{l l cccccccc}
    \hline
    \textbf{Encoder} & \textbf{Decoder} & \textbf{Acoustic (M)} & \textbf{Semantic (M)} & \textbf{Total (M)} & \textbf{WER (\%)} & \textbf{CER (\%)} & \textbf{STOI} & \textbf{PESQ} & \textbf{encRTF} \\
    \hline
    NeuCodec   & 16kHz & 35 & 600 & 635 & \textbf{2.3} & \textbf{0.9} & 0.90 & 2.06 & 0.018 \\
    NeuCodec   & 24kHz & 35 & 600 & 635 & 2.6 & 1.1 & 0.90 & 2.04 &  0.018 \\
    Distilled  & 16kHz & 21 &  21 &  42 & 2.8 & 1.2 & \textbf{0.91} & 2.11 & \textbf{0.003} \\
    Distilled  & 24kHz & 21 &  21 &  42 & 2.8 & 1.4 & 0.91 & \textbf{2.12} & \textbf{0.003} \\
    \hline
  \end{tabular}
\end{table*}

Performance is evaluated using a subset of CMU-Arctic \cite{kominek2003cmu_arctic}, where 100 utterances were randomly selected from each of the 18 speaker's data. The results are presented in Table~\ref{tab:model_performance} with a parameter breakdown of each model. Performance is measured via the Word-Error-Rate (WER) and Character-Error-Rate (CER) with transcriptions from whisper-large-v3 \cite{radford2022robustspeechrecognitionlargescale}, in addition to Short-Term Objective Intelligibility (STOI) \cite{taal2010stoi}, Perceptual Evaluation of Speech Quality (PESQ) \cite{rix2001pesq} and Real-Time-Factor of the Encoder (encRTF).

The evaluation shows that there is a limited difference in performance between the encoders when autoencoding; NeuCodec with the 16kHz decoder performs slightly better in terms of WER and CER, whereas the distilled model performs slightly better in both STOI and PESQ, which could possibly be attributed to its much larger batch-size during training. Note, the distilled encoder is also 6x faster and 15x smaller than the original encoder.

\subsection{Code-Level Analysis}
\begin{figure}[ht]
  \centering
  \includegraphics[width=.9\linewidth]{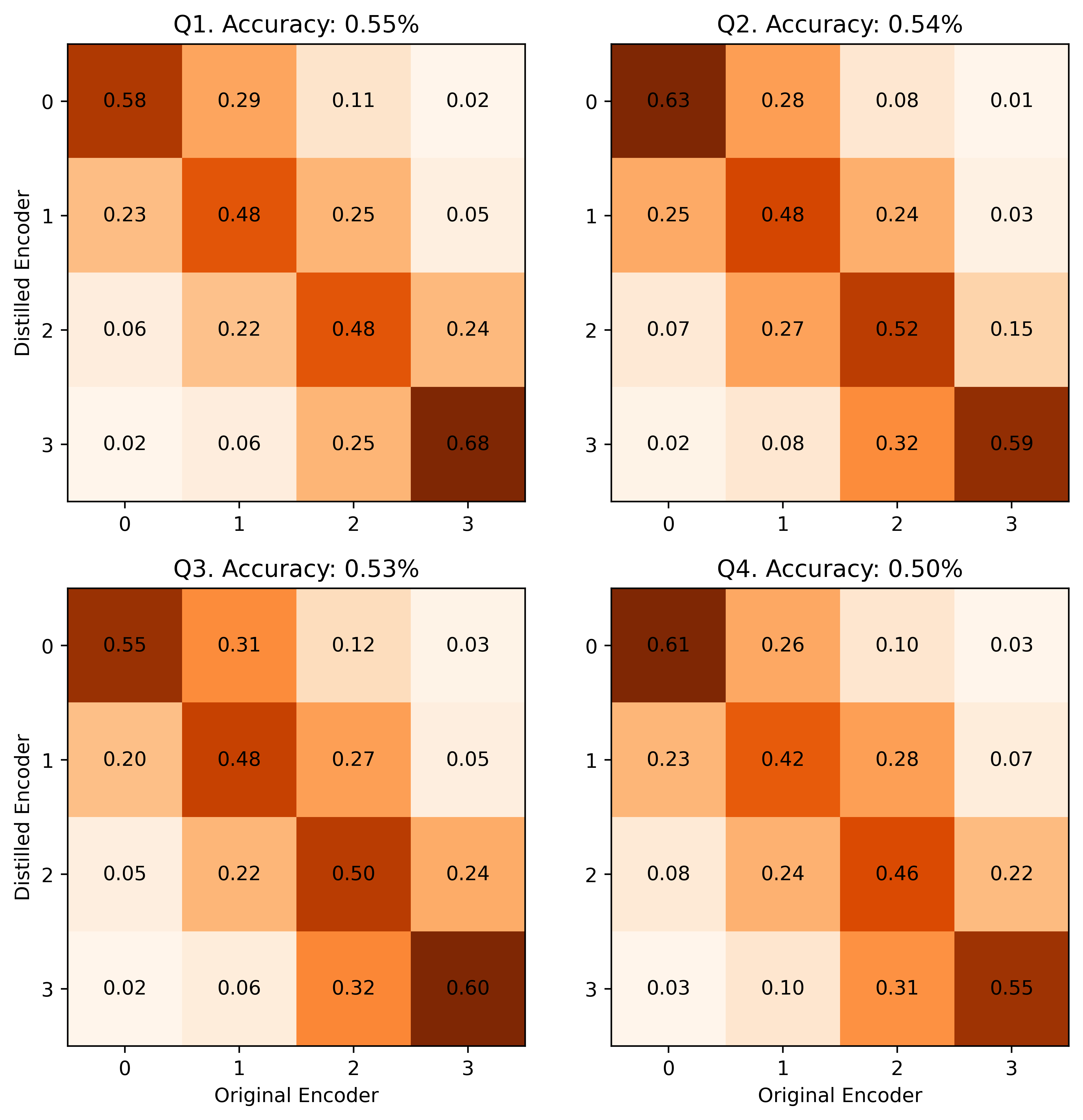}  
  \caption{A subset of implicit codebook confusion matrices between Original and Distilled Encoder level predictions. $Q_i$ refers to the index of the implicit codebook in the quantization vector.}
  \label{fig:performance_plot}
\end{figure}

Given the similar performance between the encoders, we investigate how similar their encodings are using intermediate outputs from the performance comparison. Comparing code sequences for each utterance element-wise, only 2\% of the codes match between sequences, while Mean Cosine Similarity between quantizer output projections is $0.73$. In the implicit codebook, element-wise accuracy was 53\% between sequences. A subset of codebook confusion matrices are presented in Figure~\ref{fig:performance_plot}; these show that there is reasonable incorrect classification, though notably this is primarily between neighboring levels  in each implicit codebook. As 93\% of level predictions are either correct or within a single level of the correct code, it appears that a shift by a single level is permissible for decent reconstruction and that there could be some locality to the representation in the implicit codebooks. These results show that even without learning a higher degree of code or implicit codebook correspondence, comparable reconstruction performance can still be achieved as output projections remain similar.

\section{Bit-level Perturbation Experiment}
\label{sec:bit_perturb}
When signals are transmitted through a medium, the data that is sent may be different from what is received due to signal interference or noise. How catastrophic the perturbation of a single bit is to the received signal depends on the encoding of said signal. The code indices of our quantized encoder outputs can be viewed as a bit-level digitally encoded signal, e.g. if each codebook is of size 1024 ($2^{10}$) each code index can be represented as 10 bits.

Our analysis of the output codes of our separately trained encoders shows a high level of code-level disagreement, meaning their respective bit-strings will differ substantially whilst the reconstructions will remain perceptually similar. Because there appears to be local redundancy between neighboring codes, a single perturbation in a bit-string would merely shift the code to a neighboring, perceptually similar point in the quantization space. This suggests that such perturbations would result in only limited signal degradation.

To study the robustness of the encoded sequences from both RVQ and FSQ models, we simulate the transmission of code sequences through a binary symmetric channel, where each code sequence is converted into a bit-string and each bit is transmitted incorrectly with a probability $P_{\text{flip}}$. For each codec, we encode all of Librispeech test-clean \cite{7178964}. Whilst encoding the data, we transform the integer values that correspond to individual codes in each sequence into bits using the maximum size of the codebook. Individual integer bit-strings are then concatenated into a single flat sequence, and bits are randomly perturbed (by flipping the binary value) at a given probability $P_{\text{flip}}$ across a range of values \{0.001, 0.01, 0.02, 0.05, 0.1, 0.2, 0.5\}. We then map the bit-string back to integers, reshape the flat sequence of integers into the shape of the original code sequence, and then reconstruct it.

Multiple RVQ and FSQ-based codecs are compared, as described in Table~\ref{tab:model_quantization}. StableCodec uses a modified formulation of FSQ that enables arbitrary post-hoc FSQ bottlenecks to be applied to the model; we apply a bottleneck to make the codebook a power of 2, where the quantizer levels are set to {\{8, 8, 8, 8, 4, 4\}} with a $2^{16}$ codebook size equivalent to NeuCodec that nicely fits into the bit-flipping paradigm.

Performance is measured via four metrics: STOI, PESQ, Scale-Invariant Signal-to-Distortion Ratio (SI-SDR) \cite{roux2018sdrhalfbakeddone} and Mel-Spectrogram Mean Squared Error between original and generated spectrograms.

\begin{figure}
  \centering
  \includegraphics[width=1.0\linewidth]{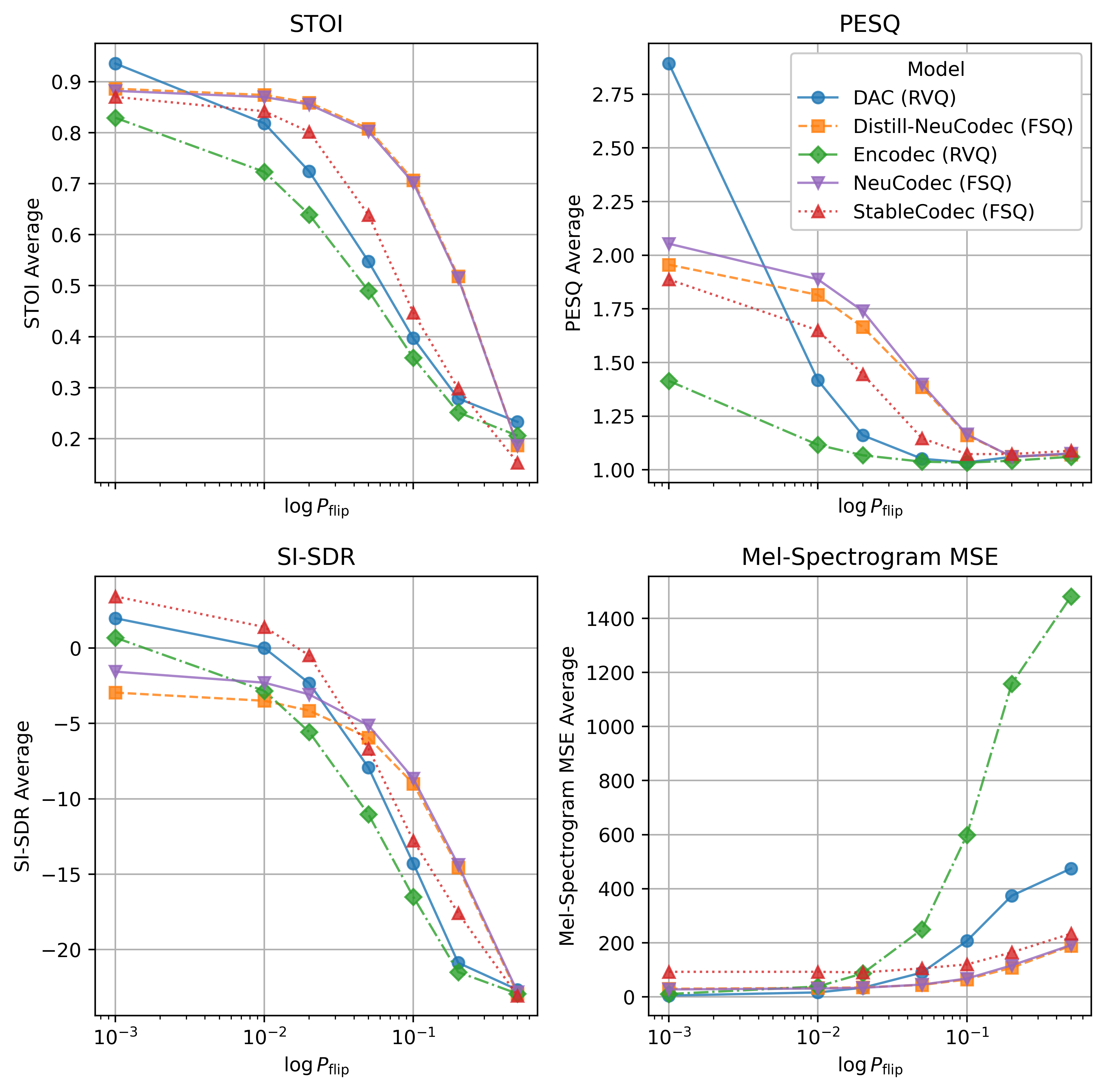}
  \vspace{-0.5cm}
  \caption{Perturbation Robustness Across conditions for all NACs.}
  \label{fig:bitflipping_results}
\end{figure}

As shown in Fig.~\ref{fig:bitflipping_results}, FSQ-based codecs maintain relatively stable performance under increasing perturbations, whereas RVQ codecs experience a sharp decline once more than 1\% of bits are altered. Notably, the STOI scores for all FSQ codecs remain high for a longer range of perturbations, indicating that - although speech quality degrades - the intelligibility remains relatively robust, even with up to 10\% of bits altered in NeuCodec.

\section{Discussion}
\label{sec:typestyle}

As shown in our distillation experiment, when encoder outputs and code sequences change, the reconstruction quality can remain the same while using same decoder. Our analysis indicates that both (1) the encoding has baked-in redundancy and (2) codes that point to different local regions in the space index similar acoustic features. Since FSQ encourages the encoder to distribute information across all codewords, as long as the codebook is large enough, redundancy becomes a feature of FSQ, as a redundant representation will be created as information spreads into all codewords regardless of the actual dimensionality of the data. Additionally, even when intentionally perturbing the code sequences, the reconstruction quality remains high compared to RVQ codecs. With FSQ, perturbations in the code indices will result in predictable size changes in embedding space. In contrast, other methods of vector quantization impose no such constraints, hence perturbing their code indices can result in arbitrarily-sized changes in the embedding space. These aspects of FSQ result in a robust method of quantization with inherent redundancy and locality in representation space.
 
\section{Conclusion}
\label{sec:typestyle}

In conclusion, we found that FSQ biases NACs to learn discrete audio encodings that have in-built redundancy and a code-level perturbation robustness that could be advantageous for designing futre low bit-rate neural compressors that are resilient to noise in transmission. Future work should assess (1) the usefulness of this property in low-latency FSQ codecs aimed at widespread deployment in transmission use-cases and (2) if the formulation of FSQ can be altered to either improve robustness properties further or to allow for direct controllability of the extent of redundancy.

\begin{table}
  \centering
  \caption{Compared FSQ and RVQ Models.}
  \label{tab:model_quantization}
  \begin{tabular}{lccc}
    \hline
    \textbf{Model} & \textbf{Rate (kHz)} & \textbf{Quantizer} & \textbf{Codebooks} \\
    \hline
    NeuCodec & 24 & FSQ & 1 \\
    Distill-NeuCodec & 24 & FSQ & 1 \\
    StableCodec \cite{parker2024scalingtransformerslowbitratehighquality} & 16 & FSQ & 1 \\
    DAC \cite{kumar2023highfidelityaudiocompressionimproved} & 24 & RVQ & 6 \\
    Encodec \cite{défossez2022highfidelityneuralaudio} & 24 & RVQ & 12 \\
    \hline
  \end{tabular}
\end{table}

\bibliographystyle{IEEEbib}
\bibliography{refs}

\end{document}